\documentclass[twocolumn,showpacs,showkeys,preprintnumbers,prd,superscriptaddress,nofootinbib]{revtex4-1}

\usepackage{amsmath}
\usepackage{amsfonts}
\usepackage{amssymb}
\usepackage{graphicx}
\usepackage{color}
\usepackage{hyperref}
\usepackage{booktabs}
\usepackage{hyperref}
\usepackage{cleveref}
\usepackage{color}

\Crefname{equation}{Eq.}{Eqs.}
\Crefname{figure}{Fig.}{Figs.}
\Crefname{section}{Sec.}{Secs.}

\usepackage{etoolbox}
\makeatletter
\appto{\appendix}{%
  \@ifstar{\def\theequation@prefix{A.}}%
          {}%
}
\makeatother

\begin{document}

\title{Extended logotropic fluids as unified dark energy models}

\author{Kuantay Boshkayev}
\email{kuantay.boshkayev@nu.edu.kz}
\affiliation{NNLOT, Al-Farabi Kazakh National University, Al-Farabi av. 71, 050040 Almaty, Kazakhstan.}
\affiliation{Department of Physics, Nazarbayev University, Kabanbay Batyr 53, 010000 Astana, Kazakhstan.}

\author{Rocco D'Agostino}
\email{rocco.dagostino@roma2.infn.it}
\affiliation{Dipartimento di Fisica, Universit\`a degli Studi di Roma ``Tor Vergata'', Via della Ricerca Scientifica 1, I-00133, Roma, Italy.}
\affiliation{Istituto Nazionale di Fisica Nucleare (INFN), Sez. di Roma ``Tor Vergata'', Via della Ricerca Scientifica 1, I-00133, Roma, Italy.}

\author{Orlando Luongo}	
\email{orlando.luongo@lnf.infn.it}
\affiliation{NNLOT, Al-Farabi Kazakh National University, Al-Farabi av. 71, 050040 Almaty, Kazakhstan.}
\affiliation{Istituto Nazionale di Fisica Nucleare (INFN), Laboratori Nazionali di Frascati, 00044 Frascati, Italy.}
\affiliation{Scuola di Scienze e Tecnologie, Universit\`a di Camerino, 62032 Camerino, Italy.}

\begin{abstract}

We here study extended classes of logotropic fluids as \textit{unified dark energy models}.
Under the hypothesis of the Anton-Schmidt scenario, we consider the universe obeying a single fluid whose pressure evolves through a logarithmic equation of state.
This result is in analogy with crystals under isotropic stresses.
Thus, we investigate thermodynamic and dynamical consequences by integrating the speed of sound to obtain the pressure in terms of the density, leading to an extended version of the Anton-Schmidt cosmic fluids.
Within this picture, we get significant outcomes expanding the Anton-Schmidt pressure in the infrared regime. The low-energy case becomes relevant for the universe to accelerate without any cosmological constant.
We therefore derive the effective representation of our fluid in terms of a Lagrangian $\mathcal{L}=\mathcal{L}(X)$, depending on the kinetic term $X$ only.
We analyze both the relativistic and non-relativistic limits.
In the non-relativistic limit we construct both the Hamiltonian and Lagrangian in terms of density $\rho$ and scalar field $\vartheta$, whereas in the relativistic case no analytical expression for the Lagrangian can be found. Thus, we obtain the potential as a function of $\rho$, under the hypothesis of irrotational perfect fluid.
We demonstrate that the model represents a natural generalization of \emph{logotropic dark energy models}. Finally, we analyze an extended class of generalized Chaplygin gas models with one extra parameter $\beta$. Interestingly, we find that the Lagrangians of this scenario and the pure logotropic one coincide in the non-relativistic regime.

\end{abstract}

\maketitle

\section{Introduction}\label{sec:intro}

The cosmological standard paradigm is currently built up in terms of pressureless matter and a positive cosmological constant \cite{Sahni00}, $\Lambda$, whose origin comes from quantum fluctuations \cite{Copeland06}. Observations making use of the corresponding  $\Lambda$CDM model provide unexpectedly small constraints over $\Lambda$, disagreeing with theoretical predictions \cite{Weinberg89}. This observational evidence jeopardizes our theoretical understanding on the standard paradigm \cite{Padmanabhan03}, leading to a severe \emph{cosmological constant problem}. Possibilities to circumvent this issue lie on abandoning $\Lambda$ in favor of a varying quintessence field \cite{Peebles88,Caldwell98} or of a dark energy contribution. Even in this case a robust physical explanation is conceivable, shifting the problem to determine which physical fluid corresponds to dark energy in the cosmic puzzle.

Among all alternatives, \emph{dark fluids} emerge as treatments which intertwine dark energy and dark matter into \emph{a single scenario}. In other words, dark energy arises from dark matter, characterizing \emph {de facto} the universe to evolve in terms of a single fluid. Dark fluids definitively represent a strategy to explore universe's dynamics without adding a new dark energy term within Einstein's equations \cite{Luongo18,Luongo14,Farnes18}. Unifying dark matter and dark energy through a single fluid is the well established as one considers Chaplygin gas \cite{Kamenshchik01}. Even though the model behaves as a pressureless fluid and a cosmological constant at early and late times respectively, it does not fulfill a suitable agreement with current data. Generalizations of Chaplygin gas have been mostly investigated \cite{Bento02,Gorini03,Bento04}, but even in this case there are severe difficulties found comparing the model with cosmic data. In Chaplygin models, a significant drawback is the net pressure generates cuspy density profiles at the center of halos in strong disagreement with observations \cite{Sandvik04} and furthermore high-redshift cosmic observations seem to be weakly compatible with Cosmic Microwave Background data.

A likely more successful unified dark fluid would overcome such caveats with a weakly increasing pressure $P$ in terms of the density $\rho$.
To this end, a logotropic version of the equation of state has been recently proposed by \cite{Chavanis15} as a natural and robust candidate for unifying dark energy and dark matter. The advantage lies on the fact that they can be obtained from first principles, i.e. they are consequence of the first principle of thermodynamics. The model provides an increasing pressure as function of $\rho$ with a logotropic temperature which turns out to be strictly positive. In turn, the corresponding dark fluid behaves as pressureless dark matter at high redshifts, whereas it shows a negative pressure at late times, pushing up the universe to accelerate. A relevant aspect of logotropic models is that they are falsifiable since they depend upon a single parameter only. The models recover the $\Lambda$CDM paradigm, breaking down before entering in the phantom regime. Moreover, logotropic dark energy prevents gravitational collapse and cusps in galaxies, overcoming the issues of Chaplygin models \cite{Chavanis17}.

Although promising scenarios, logotropic dark energy is not directly associated to a particular constituent, leaving open the challenge to understand which particles compose the logotropic fluid. In support of this fact, it has been shown that logotropic versions of dark energy fall inside a more general class based on \emph{Anton-Schmidt} fluids \cite{Anton1,Anton2}.
The Anton-Schmidt fluid empirically describes crystalline pressure for solids which deform under isotropic stress. Analogously, if one considers
the universe to deform under the action of cosmic expansion, the corresponding pressure naturally becomes
negative. This enables to model  the whole universe through a single dark counterpart. Ordinary matter, as observed in the universe, fuels the cosmic speed up as a consequence of the initial Big Bang Nucleosynthesis. Moreover, assuming a non-vanishing equation of state for matter leads to a non-pressureless matter contribution, small enough to accelerate the universe, alleviating the coincidence problem.

In this work, we show the generalization of logotropic models and we demonstrate they fall inside the picture of Anton-Schmidt fluid. To do so, we frame the evolution of the speed of sound for typical logotropic models. We thus get the most general form for the effective pressure of logotropic models. Further, we formulate both Hamiltonian and Lagrangian representations for our generalized models. Afterwards, we investigate the relativistic and non-relativistic cases, inferring the main properties derived from modifying logotropic models in featuring the universe dynamics. Last but not least, we study the equivalence between our extended logotropic models with particular Anton-Schmidt fluid. Finally, we show how the modified Chaplygin gas can be recovered from our scheme under certain conditions.

This paper is structured as follows. After this brief review on unified dark energy models, in \Cref{sec:ELM} we present a class of extended logotropic models in terms of thermodynamics quantities. Then, in \Cref{sec:lagrangian} we derive a Lagrangian formulation of the models under consideration. Finally, in \Cref{sec:conclusion} we draw the conclusions.


\section{Extended logotropic models}\label{sec:ELM}

In this section we introduce the procedure to extend logotropic models. To do so, we here assume that the universe is filled with a barotropic perfect fluid described by the Anton-Schimdt pressure \cite{Anton1}. Moreover, we simply consider the flat Friedmann-Lema\^itre-Roberston-Walker (FLRW) metric\footnote{Throughout the paper, we use units such that the speed of light is equal to unity.}:
\begin{equation}
ds^2=dt^2-a^2(t)\left(dr^2+r^2d\theta^2+r^2\sin^2\theta\ d\varphi^2\right) ,
\end{equation}
where $a(t)$ is the cosmic scale factor. Hence, the Anton-Schmidt pressure becomes
\begin{equation}
P_\text{A-S}=A\left(\dfrac{\rho}{\rho_\ast}\right)^{-\frac{1}{6}-\gamma_G}\ln\left({\dfrac{\rho}{\rho_\ast}}\right) ,
\label{eq:Anton}
\end{equation}
where $\gamma_G$ is the \emph{Gr\"uneisen parameter}  and $\rho_\ast$ is the reference density\footnote{Theoretical and observational arguments by \cite{Chavanis15} have led to identify $\rho_\ast$ with the Planck density.}. A single matter fluid obeying \Cref{eq:Anton} explains different phases of the cosmic evolution and candidates as an alternative to the standard cosmological model \cite{Anton1,Odintsov18}.

\begin{figure}[]
\begin{center}
\includegraphics[width=3.2in]{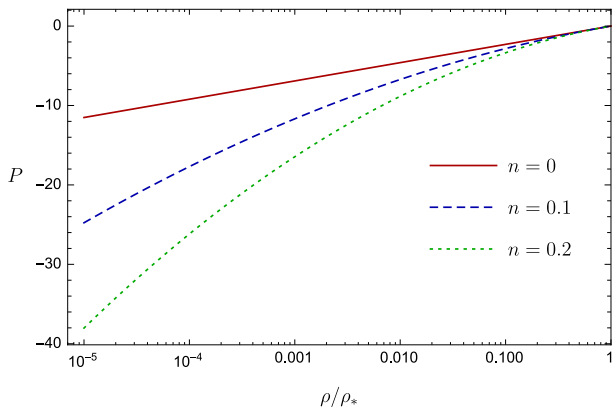}
\caption{Pressure as function of the density in extended logotropic models (cf. \Cref{P ext logotropic}). The different curves correspond to  different values of the parameter $n$, while we have assumed $A=1$.}
\label{fig:P}
\end{center}
\end{figure}

\begin{figure}
\begin{center}
\includegraphics[width=3.2in]{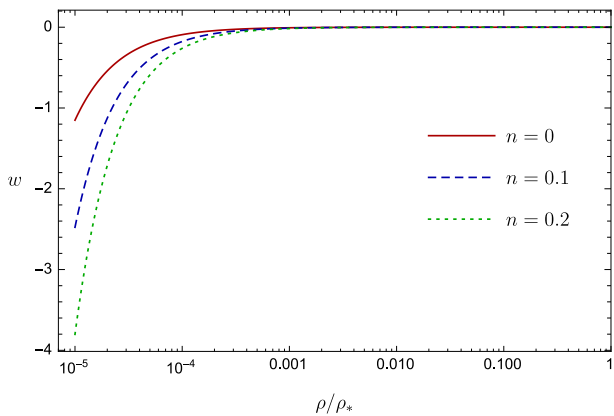}
\caption{Equation of state parameter as function of the density in extended logotropic models (cf. \Cref{w ext logotropic}). The different curves correspond to different values of the parameter $n$, while we have assumed $A/\rho_\ast=10^{-6}$.}
\label{fig:w}
\end{center}
\end{figure}

\begin{figure}
\begin{center}
\includegraphics[width=3.2in]{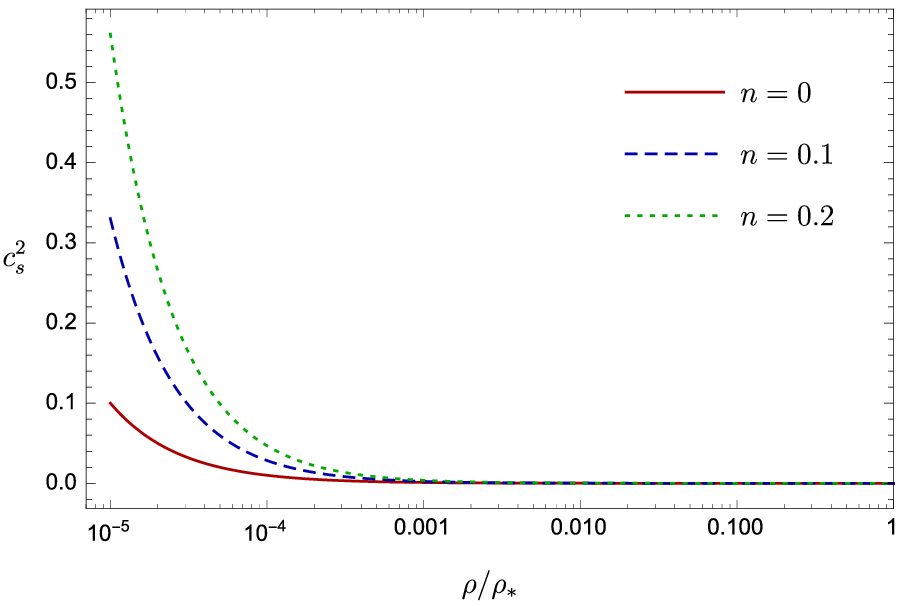}
\caption{Quadratic speed of speed as function of the density in extended logotropic models (cf. \Cref{cs ext logotropic}). The different curves correspond to different values of the parameter $n$, while we have assumed $A/\rho_\ast=10^{-6}$.}
\label{fig:cs}
\end{center}
\end{figure}

The Anton-Schmidt equation of state represents an extension of logotropic dark energy models \cite{Chavanis15}, which has been recently invoked to avoid the cosmological constant term in the Einstein field equations. In particular, the logotropic scenario is recovered in the limit $\gamma_G\rightarrow -\frac{1}{6}$. Recasting $n\equiv-\frac{1}{6}-\gamma_G$, the squared adiabatic speed of sound of the fluid with pressure given by \Cref{eq:Anton} reads
\begin{equation}
c_{s,\text{A-S}}^2\equiv \dfrac{\partial P_\text{A-S}}{\partial \rho}=\dfrac{A}{\rho}\left(\dfrac{\rho}{\rho_\ast}\right)^{-n}\left[1-n\ln\left(\dfrac{\rho}{\rho_\ast}\right)\right] .
\label{eq:cs}
\end{equation}
Thus, the corresponding equation of state is obtained by integrating \Cref{eq:cs}:
\begin{equation}
w_\text{A-S}\equiv \dfrac{P_\text{A-S}}{\rho}=\dfrac{A}{\rho}\left(\dfrac{\rho}{\rho_\ast}\right)^{-n}\ln\left({\dfrac{\rho}{\rho_\ast}}\right)+\dfrac{C}{\rho} \ ,
\end{equation}
where $C$ is an arbitrary constant that is usually assumed to be zero.
The Anton-Schmidt approach has been tested with cosmological data, which bound the parameter $\gamma_G$ to values that are compatible with $n=0$ at the 2$\sigma$ confidence level \cite{Anton1}. Motivated by these studies, we here consider an extended class of logotropic models which are obtained by expanding \Cref{eq:Anton} around $n=0$.  We thus get
\begin{equation}
P=A\left[\ln\left(\dfrac{\rho}{\rho_\ast}\right)-n\ln^2\left(\dfrac{\rho}{\rho_\ast}\right)\right] .
\label{P ext logotropic}
\end{equation}
This implies the following form for the barotropic factor:
\begin{equation}
w=\dfrac{A}{\rho}\left[\ln\left(\dfrac{\rho}{\rho_\ast}\right)-n\ln^2\left(\dfrac{\rho}{\rho_\ast}\right)\right]  .
\label{w ext logotropic}
\end{equation}
When $n=0$, the above equations recover the pure logotropic model. The speed of sound is then given by
\begin{equation}
c_s^2\equiv \dfrac{\partial P}{\partial \rho}=\dfrac{A}{\rho}\left[1-2n\ln\left(\dfrac{\rho}{\rho_\ast}\right)\right].
\label{cs ext logotropic}
\end{equation}
In \Cref{fig:P,fig:w,fig:cs}, we display the functional behaviors of pressure, equation of state and speed of sound for our model, respectively.

In the next section, we shall derive the Lagrangian formulation of such extended logotropic models.

\section{Effective field formalism}
\label{sec:lagrangian}

To proceed with the effective Lagrangian formulation of our extended logotropic models, we work in analogy with the Chaplygin case. In fact, as for the generalized Chaplygin gas \cite{Bento02}, several unified dark energy models proposed in the literature have been found to describe k-essence theories with real scalar field Lagrangians \cite{Armendariz99,Chiba99,Bilic02,Scherrer04}.
\noindent We thus consider a k-essence Lagrangian density $\mathcal{L}=\mathcal{L}(X) $, where  $X\equiv \frac{1}{2}\nabla_\mu\phi \nabla^\mu\phi$ is the kinetic term and $\phi$ is a canonical scalar field.
The energy-momentum tensor for a perfect fluid reads $T_{\mu\nu}=(\rho+P)u_\mu u_\nu-Pg_{\mu\nu}$, where $g_{\mu\nu}$ is the metric tensor, and $u_{\mu}$ is the four-velocity of the fluid given by
\begin{equation}
u_\mu=\dfrac{\nabla_\mu\phi}{\sqrt{2X}}\ .
\end{equation}
Moreover, the pressure and density of the fluid take the forms
\begin{align}
&P=\mathcal{L}(X)\ , \label{eq:P}\\
&\rho=2X\dfrac{\partial P}{\partial X}-P \label{eq:rho}\ ,
\end{align}
respectively. In order to get an effective field theory scenario for extended logotropic models, we distinguish the relativistic and the non-relativistic cases. In the next subsections, we show these cases in detail.


\subsection{Relativistic regime}

The relativistic limit over the above Lagrangian can be obtained by comparing \Cref{P ext logotropic} with \Cref{eq:P}:
\begin{equation}
\mathcal{L}=A\left[\ln\left(\dfrac{\rho}{\rho_\ast}\right)-n\ln^2\left(\dfrac{\rho}{\rho_\ast}\right)\right] .
\end{equation}
From \Cref{P ext logotropic} one also has  $\rho=\rho_\ast e^{\xi_{\pm}}$, where
\begin{equation}
\xi_{\pm}\equiv\dfrac{1}{2n}\left(1\pm\sqrt{1-\dfrac{4nP}{A}}\right).
\end{equation}
The logotropic limit is accounted as
\begin{equation}
\lim_{n\rightarrow 0^{\mp}} \xi_{\pm}= \dfrac{P}{A}\ ,
\end{equation}
leading to $\rho=\rho_\ast e^{P/A}$, consistently with what one would get from the logotropic pressure.
On the other hand, \Cref{eq:rho} holds true and implies
\begin{equation}
\int\dfrac{dX}{X}=2\int\dfrac{dP}{P+\rho_\ast e^{\xi_\pm}}\ ,
\label{eq:rel}
\end{equation}
The integral \eqref{eq:rel} on the right-hand side does not have an analytical solution. Even in the pure logotropic case $(n\rightarrow 0)$, \Cref{eq:rel} cannot be solved analytically and only numerical integration is possible.


\subsection{Non-relativistic regime}

To derive the non-relativistic Lagrangian of the extended logotropic models, we consider the classical formulation of an irrotational perfect fluid. For a given potential $V$ and for a scalar field $\vartheta$, the Hamiltonian reads
\begin{equation}
H(\rho,\vartheta,t)=\int d^3x\ \mathcal{H}=\int d^3x\left(\dfrac{1}{2}\rho\ \partial_i\vartheta\partial^i\vartheta+V(\rho)\right) ,
\label{eq:hamiltonian}
\end{equation}
where the Hamiltonian density is defined as
\begin{equation}
\mathcal{H}(\rho,\vartheta,t,x^i)=\dot{\rho}\vartheta-\mathcal{L}(\rho,\dot{\rho},t,x^i)\ .
\label{eq:hamiltonian density}
\end{equation}
Comparing \Cref{eq:hamiltonian,eq:hamiltonian density}, one finds
\begin{equation}
\mathcal{L}(\rho,\dot{\rho},t,x^i)=\dot{\rho}\vartheta-\dfrac{1}{2}\rho\ \partial_i\vartheta\partial^i\vartheta-V(\rho)\ ,
\label{eq:lagrangian}
\end{equation}
where $\rho$ and $\vartheta$ are canonically conjugate variables satisfying the Poisson bracket:
 \begin{equation}
\{\vartheta(x_i),\rho(x_j)\}=\delta(x_i-x_j)\ .
\end{equation}
Moreover, one has
\begin{align}
&\vartheta=\dfrac{\partial\mathcal{L}}{\partial\dot{\rho}}\ ,\\
&\dot{\vartheta}=\dfrac{\partial\mathcal{L}}{\partial\rho}=-\dfrac{1}{2}\partial_i\vartheta\partial^i\vartheta-V'(\rho)\ , \label{eq:dot_pi}
\end{align}
where $V'(\rho)\equiv\dfrac{\partial V}{\partial \rho}$.
In the non-relativistic scenario, the Euler equation for an ideal fluid is given by
\begin{equation}
\dot{\bm{u}}+\bm{u}\cdot \nabla\bm{u}=\bm{f}\ .
\end{equation}
In the case of isentropic motion, we have $\bm{f}=-\nabla V'(\rho)$, where $V'(\rho)$ represents the enthalpy. Furthermore, for an irrotational fluid, $\bm{u}=\nabla\vartheta$  \cite{Bazeia98}. One then has
\begin{equation}
P=\rho V'(\rho)-V\ .
\end{equation}
Taking into account \Cref{P ext logotropic}, the above equation for the pressure can be integrated into
\begin{equation}
V(\rho)=A\left[(2n-1)\left(1+\ln\left(\dfrac{\rho}{\rho_\ast}\right)\right)+n\ln^2\left(\dfrac{\rho}{\rho_\ast}\right)\right] ,
\end{equation}
where we have assumed the integration constant to be zero. One thus finds
\begin{equation}
V'(\rho)=\dfrac{A}{\rho}\left[-1+2n\left(1+\ln\left(\dfrac{\rho}{\rho_\ast}\right)\right)\right] .
\label{eq:eta}
\end{equation}
In principle, one could use the expression for $V'(\rho)$ to obtain $\rho$ from \Cref{eq:dot_pi}, and then substitute the result into \Cref{eq:lagrangian} to find the Lagrangian.
Unfortunately, for $V'(\rho)$ as given in \Cref{eq:eta}, this procedure cannot be performed analytically and thus there does not exist an explicit formula for the Lagrangian of the model with pressure (\ref{P ext logotropic}).
Nevertheless, it is possible to obtain an analytical form for the Lagrangian in the limit of pure logotropic model. For $n=0$, \Cref{eq:eta} in fact reads
\begin{equation}
V'(\rho)_{log}=-\dfrac{A}{\rho}\ ,
\end{equation}
which can be plugged into \Cref{eq:dot_pi} to obtain
\begin{equation}
\rho_{log}=\dfrac{2A}{2\dot{\vartheta}+\partial_i\vartheta\partial^i\vartheta}\ .
\end{equation}
Therefore, using \Cref{eq:P} and the expression for the logotropic pressure,
\begin{equation}
P_{log}=A\ln\left(\dfrac{\rho}{\rho_\ast}\right) ,
\label{P logotropic}
\end{equation}
one immediately finds
\begin{equation}
\mathcal{L}_{log}=A\left[\ln\left(\dfrac{A}{\rho_\ast}\right)-\ln\left(\dot{\vartheta}+\dfrac{1}{2}\partial_i\vartheta\partial^i\vartheta\right)\right] .
\label{eq:logotropic lagrangian}
\end{equation}
The above expression refers to as the Lagrangian of extended logotropic models, derived passing through the definition of Anton-Schmidt cosmic fluid. This may be interpreted as a way to relate the two approaches, i.e. matching logotropic models with Anton-Schmidt fluid. In the next section, we discuss the limit to the modified Chaplygin gas.


\section{Comparison with Chaplygin gas}

It is interesting to compare our results with the extended family of generalized Chaplygin gas models investigated by \cite{Ferreira18}.
In particular, one can consider the following k-essence Lagrangian for a perfect fluid:
\begin{equation}
\mathcal{L}=-\tilde{\rho}\left[1-(2X)^\beta\right]^{\frac{\alpha}{1+\alpha}}\ ,
\end{equation}
where $0\leq 2X\leq 1$, $\alpha$ and $\beta$ are positive constants, and $\tilde{\rho}$ is a positive constant energy density. This model is relevant since it represents a one-parameter extension of the Lagrangians proposed in the literature to study generalized Chaplygin gas \cite{Banerjee07,Beca07}.
From \Cref{eq:rho,eq:P}, one obtains
\begin{equation}
\rho=\tilde{\rho}\left(-\dfrac{P}{\tilde{\rho}}\right)^{-\frac{1}{\alpha}}\left\{1+\left(\dfrac{2\alpha\beta}{1+\alpha}-1\right)\left[1-\left(-\dfrac{P}{\tilde{\rho}}\right)^{\frac{1+\alpha}{\alpha}}\right]\right\} .
\label{eq:rho ext GCG}
\end{equation}
It is easy to verify that, for the particular choice $\beta=(1+\alpha)/2\alpha$, \Cref{eq:rho ext GCG} reduces to the generalized Chaplygin gas equation of state \cite{Bento02}:
\begin{equation}
P_{Chap}=-\dfrac{B}{\rho^\alpha}\ ,
\label{eq:P ext GCG}
\end{equation}
where $B\equiv \tilde{\rho}^{1+\alpha}$. Hence, the  speed of sound is given by
\begin{equation}
c_{s, Chap}^2\equiv\dfrac{\partial P_{Chap}}{\partial\rho}=\dfrac{\alpha B}{\rho^{1+\alpha}} \ ,
\label{eq:cs ext GCG}
\end{equation}
which is positive and subluminal if $0\leq\alpha\leq1$. Integrating \Cref{eq:cs ext GCG}, we obtain the equation of state:
\begin{equation}
w_{Chap}\equiv\dfrac{P_{Chap}}{\rho}=-\dfrac{B}{\rho^{1+\alpha}}+\dfrac{D}{\rho}\ ,
\end{equation}
with $D$ being an integration constant.
Also, one may rewrite \Cref{eq:P ext GCG} as
\begin{equation}
P_{Chap}=-\dfrac{B}{\rho_\ast^\alpha}\left(\dfrac{\rho_\ast}{\rho}\right)^\alpha+D\ ,
\end{equation}
where $B/\rho_\ast^\alpha=\tilde{\rho}$. We then expand the above expression around $\alpha=0$ to obtain
\begin{equation}
P=\tilde{\rho}\left[-1+\alpha\ln\left(\dfrac{\rho}{\rho_\ast}\right)\right]+D\ .
\label{eq:exp P GCG}
\end{equation}
Setting $D=\tilde{\rho}$ and considering the limit
\begin{equation}
\mathcal{A}=\lim_{\substack{\alpha\rightarrow 0\\ \tilde{\rho}\rightarrow\infty}} \alpha \tilde{\rho}\ ,
\end{equation}
we can finally recast \Cref{eq:exp P GCG} as
\begin{equation}
P=\mathcal{A}\ln\left(\dfrac{\rho}{\rho_\ast}\right) .
\end{equation}
We note that this expression takes the same form as the logotropic pressure given in \Cref{P logotropic}.
Therefore, adopting a similar procedure as shown in the previous section leads to the following Lagrangian in the non-relativistic regime:
\begin{equation}
\mathcal{L}=\mathcal{A}\left[\ln\left(\dfrac{\mathcal{A}}{\rho_\ast}\right)-\ln\left(\dot{\vartheta}+\dfrac{1}{2}\partial_i\vartheta\partial^i\vartheta\right)\right] ,
\end{equation}
which resembles the expression obtained in \Cref{eq:logotropic lagrangian}.


\section{Final outlooks}
\label{sec:conclusion}

In this paper, we studied an extended class of logotropic fluids as alternative scenarios to explain the current acceleration of the universe. In particular, an effective unification of dark matter and dark energy is possible in terms of a single perfect fluid whose equation of state deviates from the standard cosmological paradigm. This approach permits a natural explanation of the universe evolution without the need of \textit{ad hoc} terms in the energy-momentum tensor. In analogy to isotropic deformations of crystalline solids, we considered matter obeying the Anton-Schmidt equation of state to describe the universe deforming under the effect of cosmic expansion. The only contribution of pressureless matter with such a property is able to accelerate the universe and avoid the cosmological constant. The Anton-Schmidt approach is a generalization of the logotropic dark energy models recently proposed to unify the dark counterparts of the cosmic fluid. Specifically, the logotropic pressure is recovered from the Anton-Schmidt equation of state in the limit $n\rightarrow 0$. We thus derived a Lagrangian formulation of the models under study. Motivated by the results of observational tests, we expanded the Anton-Schmidt pressure around $n=0$ and computed the barotropic factor and the adiabatic speed of sound  for the extended logotropic model.  Assuming a homogeneous and isotropic universe, we considered the k-essence Lagrangian for a canonical scalar field. In doing so, we related the energy density and pressure to the Lagrangian density and the kinetic term. We showed that, in the relativistic regime, no analytical expression for the Lagrangian can be found. Hence, we devoted our attention to the non-relativistic regime by considering an irrotational perfect fluid with a potential $V(\rho)$ and a scalar field $\vartheta$. We thus expressed the Hamiltonian and Lagrangian densities in terms of the conjugate variables $\{\rho,\vartheta\}$. Assuming an isentropic fluid motion, we obtained the potential as a function of the density. We showed that it is possible to find an analytical form of the Lagrangian in the pure logotropic limit. Furthermore, we compared our results with the case of Chaplygin gas. To do that, we analyzed the k-essence Lagrangian of a one-parameter extension of the generalized Chaplygin gas. We thus showed that the corresponding equation of state reduces to the one of the generalized Chaplygin gas model for a particular choice of the extra parameter $\beta$. Through a suitable recasting and a series expansion around $\alpha=0$, we were able to express the pressure in the same form as in the logotropic case. Therefore, we showed that the two approaches are characterized by equivalent Lagrangian densities.


\begin{acknowledgements}

The work has been partially supported by Nazarbayev University Faculty Development Competitive Research Grants: `Quantum gravity from outer space and the search for new extreme astrophysical phenomena', Grant No. 090118FD5348 and by the MES of the RK, Program `Center of Excellence for Fundamental and Applied Physics' IRN: BR05236454, and by the MES Program IRN: BR05236494.

\end{acknowledgements}

\end{document}